\pdfoutput=1

\documentclass{moriond}

\newcommand{\Photo}{}

\usepackage{amsmath} 
\usepackage{amssymb}
\usepackage{amsfonts}
\usepackage{upgreek} 

\usepackage{ifthen} 
\newboolean{uprightparticles}
\setboolean{uprightparticles}{false} 
\usepackage{xspace} 
\usepackage{upgreek}

\def\lhcb {\mbox{LHCb}\xspace}

\def\MagUp {\mbox{\em Mag\kern -0.05em Up}\xspace}

\ifthenelse{\boolean{uprightparticles}}%
{

 \def\Ppi         {\ensuremath{\uppi}\xspace}

 \def\Ppsi        {\ensuremath{\uppsi}\xspace}

 \def\PDelta      {\ensuremath{\Delta}\xspace}                 
 \def\PXi      {\ensuremath{\Xi}\xspace}                 
 \def\PLambda      {\ensuremath{\Lambda}\xspace}                 
 \def\PSigma      {\ensuremath{\Sigma}\xspace}                 
 \def\POmega      {\ensuremath{\Omega}\xspace}                 
 \def\PUpsilon      {\ensuremath{\Upsilon}\xspace}                 
 
 \def\PB      {\ensuremath{\mathrm{B}}\xspace}                 
                  
 \def\PD      {\ensuremath{\mathrm{D}}\xspace}

 \def\PJ      {\ensuremath{\mathrm{J}}\xspace}                 
 \def\PK      {\ensuremath{\mathrm{K}}\xspace}

 \def\Pb      {\ensuremath{\mathrm{b}}\xspace}                 
 \def\Pc      {\ensuremath{\mathrm{c}}\xspace}                 
                  
 \def\Pe      {\ensuremath{\mathrm{e}}\xspace}

 \def\Pi      {\ensuremath{\mathrm{i}}\xspace}

 \def\Pp      {\ensuremath{\mathrm{p}}\xspace}

}
{

 \def\Ppi         {\ensuremath{\pi}\xspace}

 \def\Ppsi        {\ensuremath{\psi}\xspace}                 
                  
 \mathchardef\PDelta="7101
 \mathchardef\PXi="7104
 \mathchardef\PLambda="7103
 \mathchardef\PSigma="7106
 \mathchardef\POmega="710A
 \mathchardef\PUpsilon="7107
                  
 \def\PB      {\ensuremath{B}\xspace}                 
                  
 \def\PD      {\ensuremath{D}\xspace}

 \def\PJ      {\ensuremath{J}\xspace}                 
 \def\PK      {\ensuremath{K}\xspace}

 \def\Pb      {\ensuremath{b}\xspace}                 
 \def\Pc      {\ensuremath{c}\xspace}                 
                  
 \def\Pe      {\ensuremath{e}\xspace}

 \def\Pi      {\ensuremath{i}\xspace}

 \def\Pp      {\ensuremath{p}\xspace}

}

\makeatletter
\ifcase \@ptsize \relax
  \newcommand{\miniscule}{\@setfontsize\miniscule{4}{5}}
\or
  \newcommand{\miniscule}{\@setfontsize\miniscule{5}{6}}
\or
  \newcommand{\miniscule}{\@setfontsize\miniscule{5}{6}}
\fi
\makeatother

\DeclareRobustCommand{\optbar}[1]{\shortstack{{\miniscule (\rule[.5ex]{1.25em}{.18mm})}
  \\ [-.7ex] $#1$}}

\def\en         {{\ensuremath{\Pe^-}}\xspace}   
\def\ep         {{\ensuremath{\Pe^+}}\xspace}

\def\cquark    {{\ensuremath{\Pc}}\xspace}

\def\bquark    {{\ensuremath{\Pb}}\xspace}

\def\pion   {{\ensuremath{\Ppi}}\xspace}

\def\pip    {{\ensuremath{\pion^+}}\xspace}
\def\pim    {{\ensuremath{\pion^-}}\xspace}
\def\pipm   {{\ensuremath{\pion^\pm}}\xspace}

\def\kaon    {{\ensuremath{\PK}}\xspace}
  \def\Kbar    {{\kern 0.2em\overline{\kern -0.2em \PK}{}}\xspace}

\def\KorKbar    {\kern 0.18em\optbar{\kern -0.18em K}{}\xspace}

\def\Kp      {{\ensuremath{\kaon^+}}\xspace}
\def\Km      {{\ensuremath{\kaon^-}}\xspace}

\def\Kmp     {{\ensuremath{\kaon^\mp}}\xspace}
\def\KS      {{\ensuremath{\kaon^0_{\mathrm{ \scriptscriptstyle S}}}}\xspace}

  \def\Dbar    {{\kern 0.2em\overline{\kern -0.2em \PD}{}}\xspace}
\def\D       {{\ensuremath{\PD}}\xspace}

\def\DorDbar    {\kern 0.18em\optbar{\kern -0.18em D}{}\xspace}
\def\Dz      {{\ensuremath{\D^0}}\xspace}

\def\Dstarpm {{\ensuremath{\D^{*\pm}}}\xspace}

\def\Bbar    {{\ensuremath{\kern 0.18em\overline{\kern -0.18em \PB}{}}}\xspace}

\def\BorBbar    {\kern 0.18em\optbar{\kern -0.18em B}{}\xspace}

\def\jpsi     {{\ensuremath{{\PJ\mskip -3mu/\mskip -2mu\Ppsi\mskip 2mu}}}\xspace}

  \def\Y#1S{\ensuremath{\PUpsilon{(#1S)}}\xspace}

\def\proton      {{\ensuremath{\Pp}}\xspace}
\def\antiproton  {{\ensuremath{\overline \proton}}\xspace}

\def\Lz          {{\ensuremath{\PLambda}}\xspace}
\def\Lbar        {{\ensuremath{\kern 0.1em\overline{\kern -0.1em\PLambda}}}\xspace}
\def\LorLbar    {\kern 0.18em\optbar{\kern -0.18em \PLambda}{}\xspace}

\def\to                 {\ensuremath{\rightarrow}\xspace}

\def\AT#1     {\ensuremath{A_{\mathrm{T}}^{#1}}\xspace}           

\def\C#1      {\ensuremath{\mathcal{C}_{#1}}\xspace}                       
\def\Cp#1     {\ensuremath{\mathcal{C}_{#1}^{'}}\xspace}                    
\def\Ceff#1   {\ensuremath{\mathcal{C}_{#1}^{\mathrm{(eff)}}}\xspace}        
\def\Cpeff#1  {\ensuremath{\mathcal{C}_{#1}^{'\mathrm{(eff)}}}\xspace}       
\def\Ope#1    {\ensuremath{\mathcal{O}_{#1}}\xspace}                       
\def\Opep#1   {\ensuremath{\mathcal{O}_{#1}^{'}}\xspace}                    

\newcommand{\tev}{\ifthenelse{\boolean{inbibliography}}{\ensuremath{~T\kern -0.05em eV}}{\ensuremath{\mathrm{\,Te\kern -0.1em V}}}\xspace}
\newcommand{\gev}{\ensuremath{\mathrm{\,Ge\kern -0.1em V}}\xspace}
\newcommand{\mev}{\ensuremath{\mathrm{\,Me\kern -0.1em V}}\xspace}
\newcommand{\kev}{\ensuremath{\mathrm{\,ke\kern -0.1em V}}\xspace}
\newcommand{\ev}{\ensuremath{\mathrm{\,e\kern -0.1em V}}\xspace}
\newcommand{\gevc}{\ensuremath{{\mathrm{\,Ge\kern -0.1em V\!/}c}}\xspace}
\newcommand{\mevc}{\ensuremath{{\mathrm{\,Me\kern -0.1em V\!/}c}}\xspace}
\newcommand{\gevcc}{\ensuremath{{\mathrm{\,Ge\kern -0.1em V\!/}c^2}}\xspace}
\newcommand{\gevgevcccc}{\ensuremath{{\mathrm{\,Ge\kern -0.1em V^2\!/}c^4}}\xspace}
\newcommand{\mevcc}{\ensuremath{{\mathrm{\,Me\kern -0.1em V\!/}c^2}}\xspace}

\def\mum  {\ensuremath{{\,\upmu\mathrm{m}}}\xspace}

\def\invnb {\ensuremath{\mbox{\,nb}^{-1}}\xspace}

\def\gsim{{~\raise.15em\hbox{$>$}\kern-.85em
          \lower.35em\hbox{$\sim$}~}\xspace}
\def\lsim{{~\raise.15em\hbox{$<$}\kern-.85em
          \lower.35em\hbox{$\sim$}~}\xspace}

\def\ptot       {\mbox{$p$}\xspace}
\def\pt         {\mbox{$p_{\mathrm{ T}}$}\xspace}

\def\tell1  {TELL1\xspace}
\def\ukl1   {UKL1\xspace}

\def\sqsnn   {\ensuremath{\protect\sqrt{s_{\scriptscriptstyle\rm NN}}}\xspace}
\def\pHe{\ensuremath{p\text{He}}\xspace}
\def\pp{\proton\proton}
\def\pbar{\antiproton}

\def\pe{\proton\en\xspace}

\def\pvz{\ensuremath{ z}\xspace}

\def\DLLppi{\ensuremath{\mathrm{DLL (p - \pi)}}\xspace}
\def\DLLpK{\ensuremath{\mathrm{DLL (p - K)}}\xspace}

\begin{document}
\vspace*{4cm}
\title{Fixed target measurements at LHCb for cosmic rays physics}

\author{ GIACOMO GRAZIANI}

\address{INFN, Sezione di Firenze, via Sansone 1, 50019 Sesto Fiorentino (FI), Italy}

\maketitle\abstracts{
The LHCb experiment has the unique possibility, among the LHC experiments, 
to be operated in fixed target mode, using its internal gas target.
The energy scale achievable at the LHC, combined with the \lhcb
forward geometry and detector capabilities, allow to explore particle production in a
wide Bjorken-$x$ range at the $\sqsnn\sim 100$ \gev energy scale,  
providing novel inputs to nuclear and cosmic ray physics. 
The first measurement of antiproton production in collisions of LHC
protons on helium  nuclei at rest is presented. The knowledge of
this cross-section is of great importance for the study of the cosmic
antiproton flux, and the \lhcb results are expected to improve the
interpretation of the recent high-precision measurements of cosmic
antiprotons performed by the space-borne PAMELA and AMS-02 experiments.
}

\section{LHCb as a fixed target detector}
The \lhcb detector~\cite{Alves:2008zz} is a single-arm forward
spectrometer covering the \mbox{pseudorapidity} range $2<\eta <5$,
designed for the study of particles containing \bquark or \cquark
quarks, which are predominantly produced at high $\eta$ in \pp collisions
at the LHC. The forward geometry 
and excellent vertexing, tracking and particle identification (PID)
capabilities~\cite{LHCb-DP-2014-002},  which are key features for the
reconstruction of heavy flavour decays, make it also an ideal tool  
to study interactions of the LHC beams with a fixed target. Such target is provided
by the SMOG (System for Measuring Overlap with Gas) device~\cite{smog,LHCb-PAPER-2014-047}, 
through which tiny amounts of a noble gas (He,
Ne, Ar) can be injected inside the primary LHC vacuum around  the \lhcb
vertex detector (VELO). 
The design gas
pressure in the VELO region is $2 \times 10^{-7}$ mbar, which is small
enough not to significantly perturb the 
LHC operation. The device was originally conceived to determine the
machine luminosity using a beam gas imaging
technique\cite{LHCb-PAPER-2014-047}. 
Since 2015, \lhcb has started to exploit SMOG to perform a
set of physics runs, using special fills not devoted to \pp
physics, with different beam and target configurations, allowing a wealth of unique production studies.
One of the main goals of this program is the study
of heavy flavour production in proton-ion collisions with different
target mass number at $\sqsnn\sim 100$ \gev, an intermediate energy
between the existing data collected at SPS and RHIC/LHC
accelerators. These measurements can shed light on the cold nuclear
matter effects affecting the production of the most relevant probes
that are used for detecting  quark-gluon plasma at higher energy
density. 
Another attractive feature of the fixed target configuration is the
access to the large Bjorken-$x$ region in the target nucleus. Nuclear
PDFs in this region are sensitive to antishadowing effects and to
possible contributions from intrinsic charm and beauty. The first
results for charm production, obtained from 
a data set of proton-argon collisions at $\sqsnn= 110$ \gev
corresponding to a few \invnb, have been recently
released~\cite{LHCB-CONF-2017-001}. Though the results are still
limited by the data size,  the observed differential \Dz and
\jpsi yields are already expected to provide constrains on nuclear PDFs at large $x$.
Exclusive particle production studies in this kinematic range
can also provide crucial inputs to the modelling of cosmic ray showers
in the atmosphere and in the cosmos.

\section{Cosmic collisions at LHCb}

The measurement discussed in the following is motivated by the
high-precision determination of  the \antiproton/\proton ratio in
cosmic rays, up to the energy of 350 \gev, achieved during the last years by the
space-borne PAMELA~\cite{pamela} and AMS-02~\cite{ams} experiments. 
The investigation of the antimatter content in cosmic rays is recognized as a primary
tool for the understanding of high-energy astrophysical phenomena and 
the measurements of the antiproton fraction outside of the Earth's atmosphere provides
a sensitive indirect probe for Dark Matter.
The interpretation of these measurements is currently
limited by the uncertainty on the expected amount of secondary
antiprotons produced by spallation of primary cosmic rays on the
interstellar medium. State-of-the-art
calculations~\cite{diMauro:2014zea,Giesen:2015ufa,Kappl:2015bqa}
show that the experimental results are still compatible
with the secondary \pbar production, tough data indicate a larger \pbar
flux at high energy with respect to most predictions.
The largest uncertainty on the prediction 
is due, particularly in the 10--100 GeV range, to the limited knowledge
of the \antiproton production cross-section in the relevant processes.
In particular, no data for \pbar production exist for \pHe collisions.

\section{Measurement of antiproton production in \pHe collisions}

\lhcb performed the first measurement of \pbar production in \pHe
collisions by operating SMOG with helium during special fills with
limited number of proton bunches, accelerated to 6.5 TeV (\sqsnn=110 \gev).
Most of the data were collected in a single LHC fill during about five
hours in May 2016.
Events were triggered with a minimum bias requirement, fully efficient
on the collisions producing an antiproton within the detector acceptance.
The measurement is performed from collisions occurring in an 80 cm
long fiducial region, where the best reconstruction efficiency is achieved. 
Antiprotons are counted in two-dimensional bins in momentum (\ptot)
and transverse momentum (\pt), in the
range $12 < \ptot < 110$ \gevc, $0.4 < \pt < 4$ \gevc. The kinematic
limits are dictated by the acceptance of the two ring-imaging Cherenkov
detectors (RICH) providing particle identification.
The first one covers  the range $2<\eta<4.4$ and
allows \pbar/\Km separation in the momentum range 10-60 \gevc, while the second
has acceptance $3<\eta<5$ and  actively identifies antiprotons
between 30 and about 110 \gev.
The analysis described in this document covers only the prompt \pbar production,
namely the antiprotons produced directly in the \pHe collision, or
from resonances decaying via the strong interaction. The component due to hyperon decays,
treated here as a background component and subtracted from the result, will be the subject of 
a dedicated study.

\subsection{Reconstruction}


Candidates are selected by requiring a negative track in the kinematic
range of interest after applying quality requirements on the
reconstruction of the track  and of the collision primary vertex (PV),
whose position must be compatible with the beam geometry and lie
within a 80 cm long fiducial
region  where the best reconstruction efficiency is achieved. 
These requirements are almost fully efficient
in simulation, while allowing to suppress vertices 
from decays, secondary collisions, or combinatorial track association 
in events produced by collisions occurring upstream of the fiducial region. 
The PV reconstruction efficiency is estimated from
simulation. The average value varies with \pvz from 76\% in the most
upstream region to 95\% around the nominal collision point, with a
mild dependence on the \pt of the candidate \pbar.
The related systematic uncertainty is evaluated by weighting the
simulation to account for discrepancies in the PV topology description
with respect to data. 
The reconstruction efficiency for prompt antiprotons, $\epsilon_{\rm
  rec}$, including acceptance effects and the tracking detector
efficiency, is determined 
from simulation in three-dimensional bins of \ptot, \pt and \pvz. The 
bins are chosen to be small enough to minimize the dependence on the
assumed spectra in simulation, but are at least five times larger than the
resolution in each variable.
The size of the kinematic bins is chosen according to the
expected spectrum of reconstructed antiprotons, while
twelve bins of equal
size are used for \pvz. To control systematic uncertainties, only bins
where $\epsilon_{\rm rec}$ exceeds 35\% are retained. 
The tracking efficiency predicted by
the simulation, averaged over $z$, ranges from 40 to 80\%  depending on the track
kinematics. A correction determined from calibration
samples in \pp data, of order 1\%, is applied to account for
imperfections in the simulation of the tracking detector response.

\subsection{Particle identification}
\begin{figure}[tb]
    \centering
    \includegraphics[width=\textwidth]{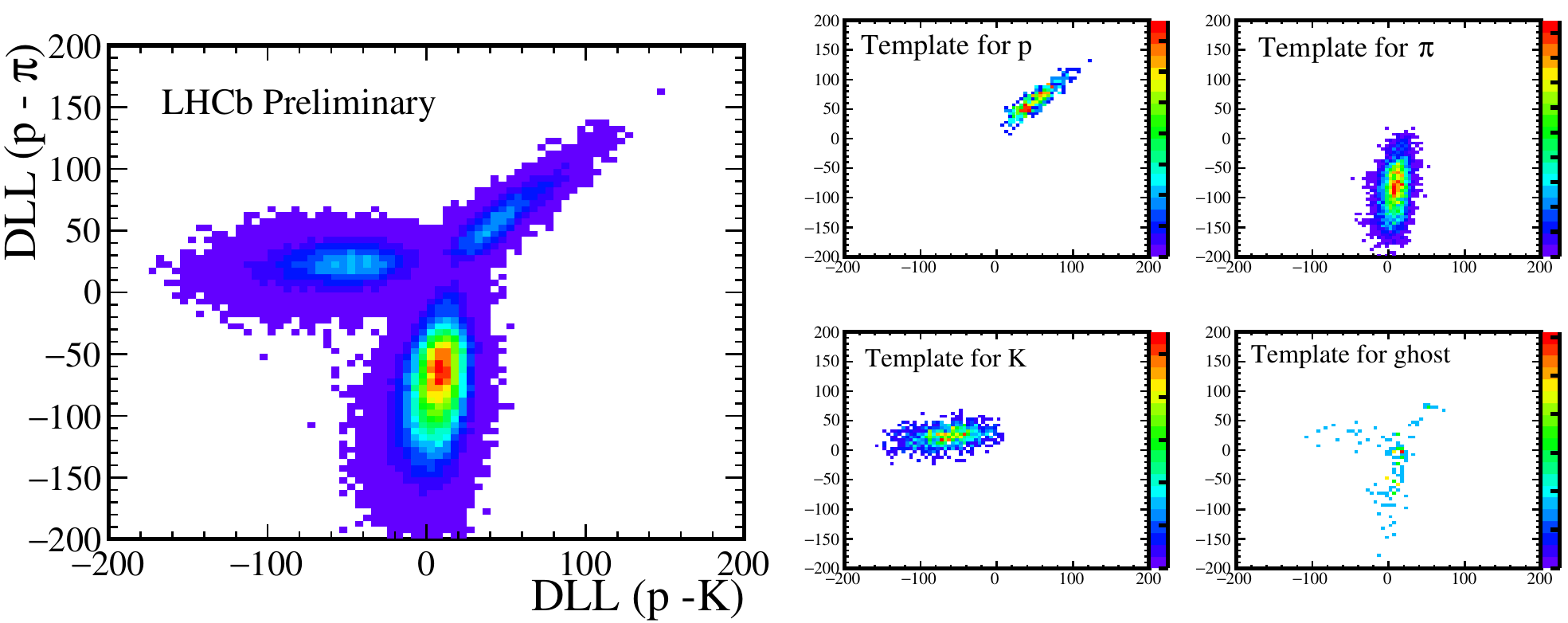}

    \caption{\label{fig:pidfitExample}
      Example of two-dimensional DLL distributions for a particular bin ($ 21.4 <
      \ptot < 24.4 \gevc, 1.2 < \pt < 1.5 \gevc$), illustrating the performance
      of the RICH detectors in separating \pim, \Km and \pbar particles. 
      The distribution in data is shown on the left, while 
      the templates for the four categories, obtained from calibration samples 
      (from simulation for the ghost component), are shown on the right.}
  \end{figure}
Antiprotons within the selected sample of negative tracks are identified through the response of the RICH
detectors, from which two variables are built, \DLLppi and \DLLpK,
representing the difference of the log likelihood between  the proton and pion 
and the proton and kaon hypothesis. 
The fraction of antiprotons among the
negative tracks is determined from the two-dimensional distribution of the
DLL variables. Three sets of templates for the different particles species are considered:
predicted by the simulation, obtained from calibration samples in the
\pHe data set, and from calibration samples in the \lhcb\xspace \pp data collected in 2016.
The \pHe calibration samples consist of selected $\KS\to\pip\pim$
decays for pions, $\Lz\to\proton\pim (\Lbar\to\antiproton\pip)$ for (anti)protons and
$\phi\to\Kp\Km$ for kaons. 
Calibration samples in \pp data have much larger size, and,
thanks to the available $\Dstarpm\to\Dz(\Kmp\pipm)\pipm$ selection, provide better 
coverage of the region with high \ptot and \pt for kaons. On the other
hand they are characterized by a much larger detector occupancy, which is critical for 
the RICH performance, with respect to the \pHe events. Such difference
is taken into account by weighting the \pp events according to the
observed charged track multiplicity.
The most appropriate calibration templates are chosen for each
kinematic region, while the related systematic uncertainty is
estimated from their comparison.
The fraction of antiprotons in each kinematic bin of the selected sample is determined with 
a two-dimensional extended binned maximum likelihood  fit, where the 
(\DLLppi, \DLLpK) distribution in data is fitted as the sum of four components: \pim, \Km, \pbar 
and ghost tracks. The latter component, whose template is obtained from
simulation,  is needed to account for candidate tracks which can not 
be unambiguously matched to a particle. This occurs in simulation in
about 2\% of cases.
The DLL distributions for data and calibration samples, illustrating
the RICH performance, are shown
in Fig.~\ref{fig:pidfitExample} for an arbitrary kinematic bin.

\subsection{Backgrounds}

The selected antiprotons that are not prompt are treated as a background 
and are subtracted from the selected sample. Such background is
suppressed by requiring that the impact parameter (IP), which is
reconstructed through the VELO with a resolution of $\left(15+29/\pt
(\gevc) \right)\mum$,  is compatible with zero.
The residual nonprompt background varies in simulation between 3\% at
the lowest \pt values and 1\% at high \pt. 
In 90\% of cases, this background is due to hyperon decays, while in the remaining cases 
the antiprotons come from a secondary collision in the detector material and
are mistakenly assigned to the primary vertex. 
The average background level is constrained from the tail of the IP distribution
in data to be
$(2.6 \pm 0.6) \%$, where the
uncertainty is systematic and is estimated by varying the fraction of
nonprompt \pbar in simulation within the range where a good agreement
with data is observed.  

\begin{figure}[tb]
    \centering
    \includegraphics[width=.86\textwidth]{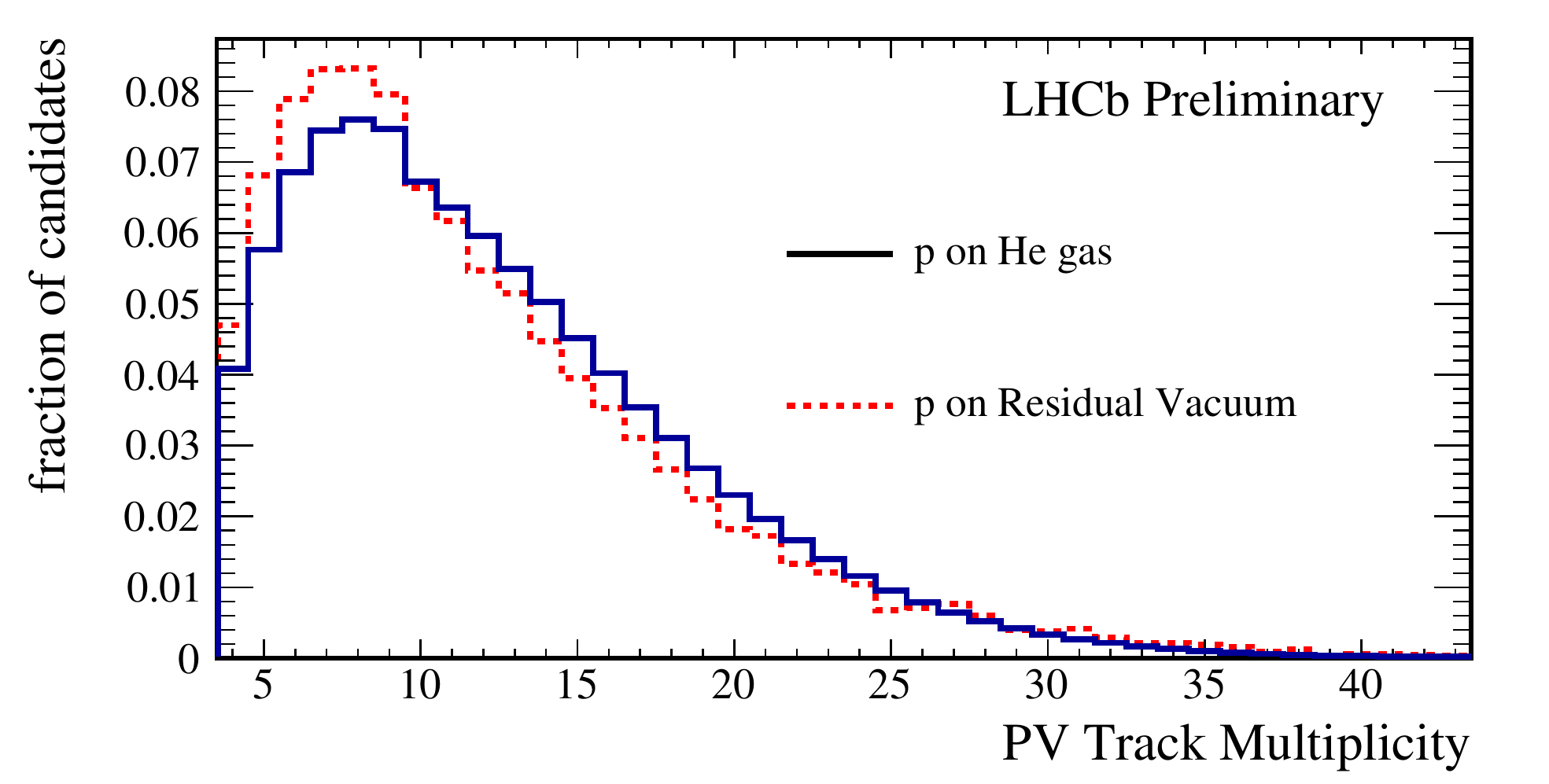}
    \caption{PV track multiplicity
      distributions for collisions
      on residual vacuum or on the helium target.}
    \label{fig:multnogas}
\end{figure}

Another background to be considered is due to the possible
contamination of the gas target.
The rate of collisions on the LHC residual vacuum is evaluated by
acquiring part of the data without the injected helium gas, while
using the same vacuum-pumping configuration of the data taking with gas.
The yield measured in these special data  are scaled according to the corresponding number of
protons on target, and  the relative average contribution from residual vacuum
is evaluated to be $(0.7 \pm 0.2) \%$, where the uncertainty is systematic. 
In Fig.~\ref{fig:multnogas},  the  normalized PV multiplicity distributions
for data with and without injected gas are compared. The
multiplicity is slightly lower on average for the collisions on the
residual gas, though it exhibits a longer tail. This confirms that
residual vacuum is dominated by hydrogen, with a small contribution
from elements heavier than helium, as indicated by the rest gas
analysis performed by the LHC vacuum group in the absence of beam. 

\subsection{Normalization}

The SMOG device does not presently allow a precise calibration
of the injected gas pressure. Instead, the normalization for
the \pbar production measurement is provided by observing a process with a well-known
cross-section. 
Single electrons scattered off by the proton beam can be observed within the \lhcb acceptance.
For a 6.5 TeV proton beam, in the corresponding kinematic range 
the scattering is purely elastic.
The cross-section in the polar angle range $3 < \theta < 27$ mrad,
outside of which the electrons can not be reconstructed in 
\lhcb, is 180.6 $\mu$b.
Though this is three orders of magnitude
below the total nuclear inelastic cross-section, events are expected
to have a distinct signature, 
with a single low-momentum and low-\pt electron track visible in the
detector, with little or no other 
activity. Background events which could mimic this signature are
expected from soft nuclear interactions 
where the candidate electron is either the product of a photon
conversion, or a light hadron from a central exclusive production
event. In both cases the background is charge symmetric. This allows
to model the background from events with a single positively
charged track (referred to as single positrons in the following). 
Multiplicity distributions in data confirm that background-dominated
regions close to the signal are charge symmetric.
\begin{figure}[tb]
  \centering
  \includegraphics[width=\textwidth]{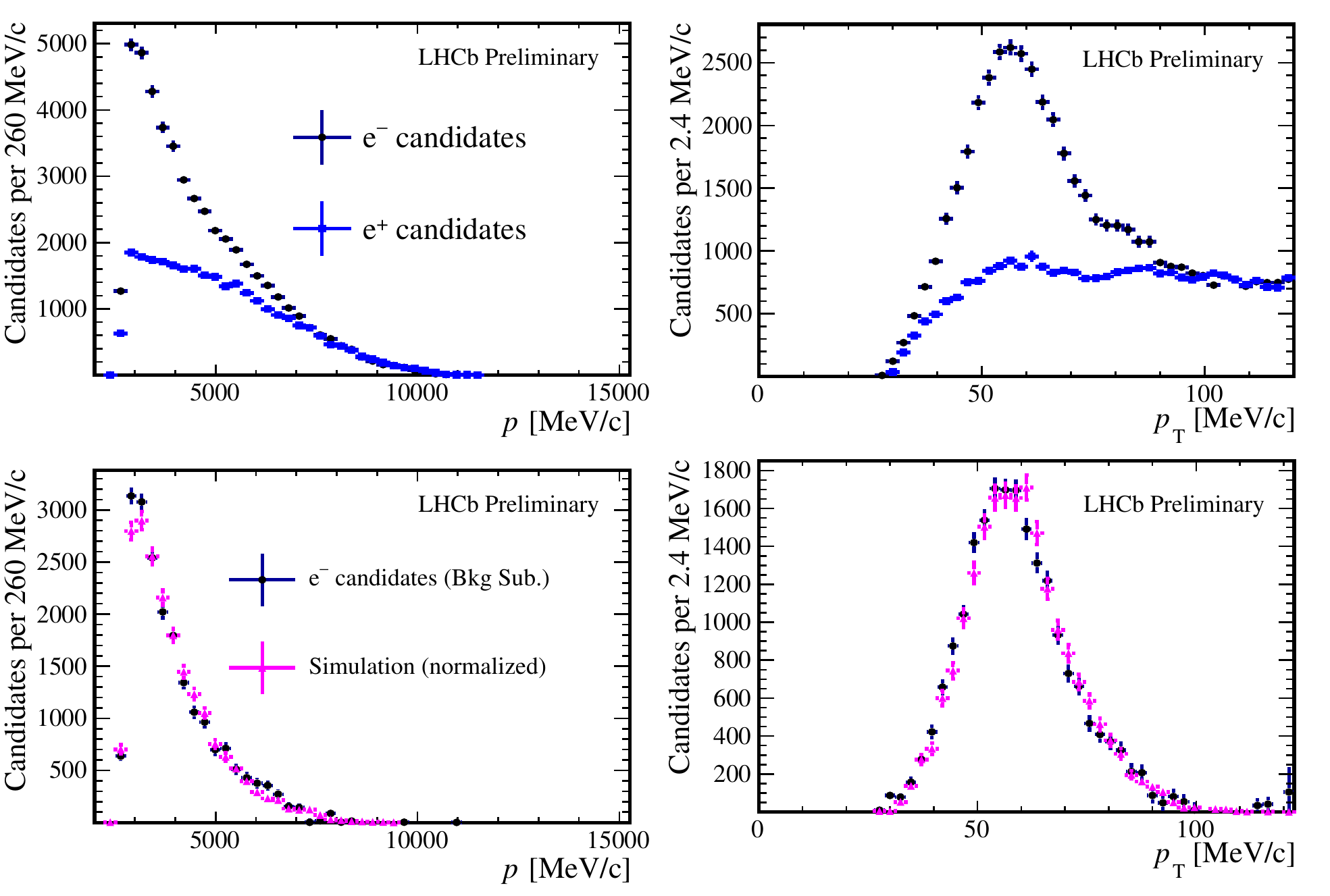}
  \caption{Distributions of (left) momentum and (right) \pt for (top plot) single electron and
    single positron candidates; (bottom plot) background subtracted electron candidates, compared with the
    distributions in simulation, which are normalized to data. }
  \label{fig:en_kin}
\end{figure}
Single electrons candidate events are selected through a loose kinematic
selection on the track and applying veto requirements on any detector activity  not
compatible with the elastic scattering hypothesis.
The selection yields 16569 single \en candidates and 9548 \ep
candidates. The signal yield is obtained by the difference of the two
components. The background subtracted  kinematic
distributions are shown on Fig.~\ref{fig:en_kin}.
An excellent agreement with the simulation is
observed, confirming the validity of the charge-symmetry hypothesis
for the background.

The luminosity is determined from the background-subtracted yield of scattered
electrons $N_e$, the known cross-section $\sigma_{\pe}$ and the electron
reconstruction efficiency $\epsilon_{\rm e}$, as 
$\mathcal{L} = N_e/ (Z_{He} \times \sigma_{\pe} \times \epsilon_{\rm e})$,
where $Z_{He}=2$ is the helium atomic number. Possible effects of gas
ionization were evaluated and are expected to be negligible.
The reconstruction efficiency, evaluated from simulation, is limited  by the soft 
momentum and transverse momentum spectrum. Electrons lose a sizable
fraction of their energy through bremsstrahlung in the beam pipe and
detector material, and large acceptance effects are caused by the
spectrometer magnetic field. The relative systematic uncertainty on
$\epsilon_{\rm e}$ is estimated to be 6\% from the stability of the
result when varying the main selection criteria, and in particular
from the ability of the simulation to describe the large modulation 
of the efficiency with the electron azimuthal angle.
The result is $\mathcal{L} = 0.443 \pm 0.011 \pm
0.027$ nb$^{-1}$, where the first uncertainty is statistical, and the
second is systematic, dominated by the uncertainty on $\epsilon_{\rm e}$.

\begin{table}[h]
  \centering
\caption{Relative uncertainties on the \pbar production cross-section measurement. 
The ranges refer to the variation among kinematic bins. }
\label{tab:syst}  
\begin{tabular}{lccc}
\hline
Source  & Statistical & Systematic  & Systematic \\ 
        &             & (correlated)  & (uncorrelated) \\ \hline
Data size        &  ~~~$0.7 - 10.8\%$~~~  & & \\
                 &    ($<3\%$ for most bins)  & & \\
Normalization      &$2.5 \%$ & $6.0 \%$ & \\
Event and PV requirements & &$0.3 \%$ & \\
PV reco                & &   $0.8 \%$ & \\
Tracking               & &   $2.2 \%$ &  $3.2 \%$ \\
Nonprompt background  &  & $0.3 - 0.7 \%$ &  \\
Residual vacuum background   & &   $0.1 \%$ & \\
Efficiency of IP requirement   & &  &   $1.0 \%$ \\
PID                    & & $1.2 - 5.0 \%$  &  $0 - 26 \%$ \\
                       & &   &   ($<10\%$ for most bins)  \\
Simulated sample size  & & &  $0.8 - 15 \%$ \\
                       & &   &($<4\%$ for $\pt<2$ \gevc)  \\
\hline
\end{tabular}
\end{table}

\begin{figure}[h]
  \centering
    \includegraphics[width=.9\textwidth]{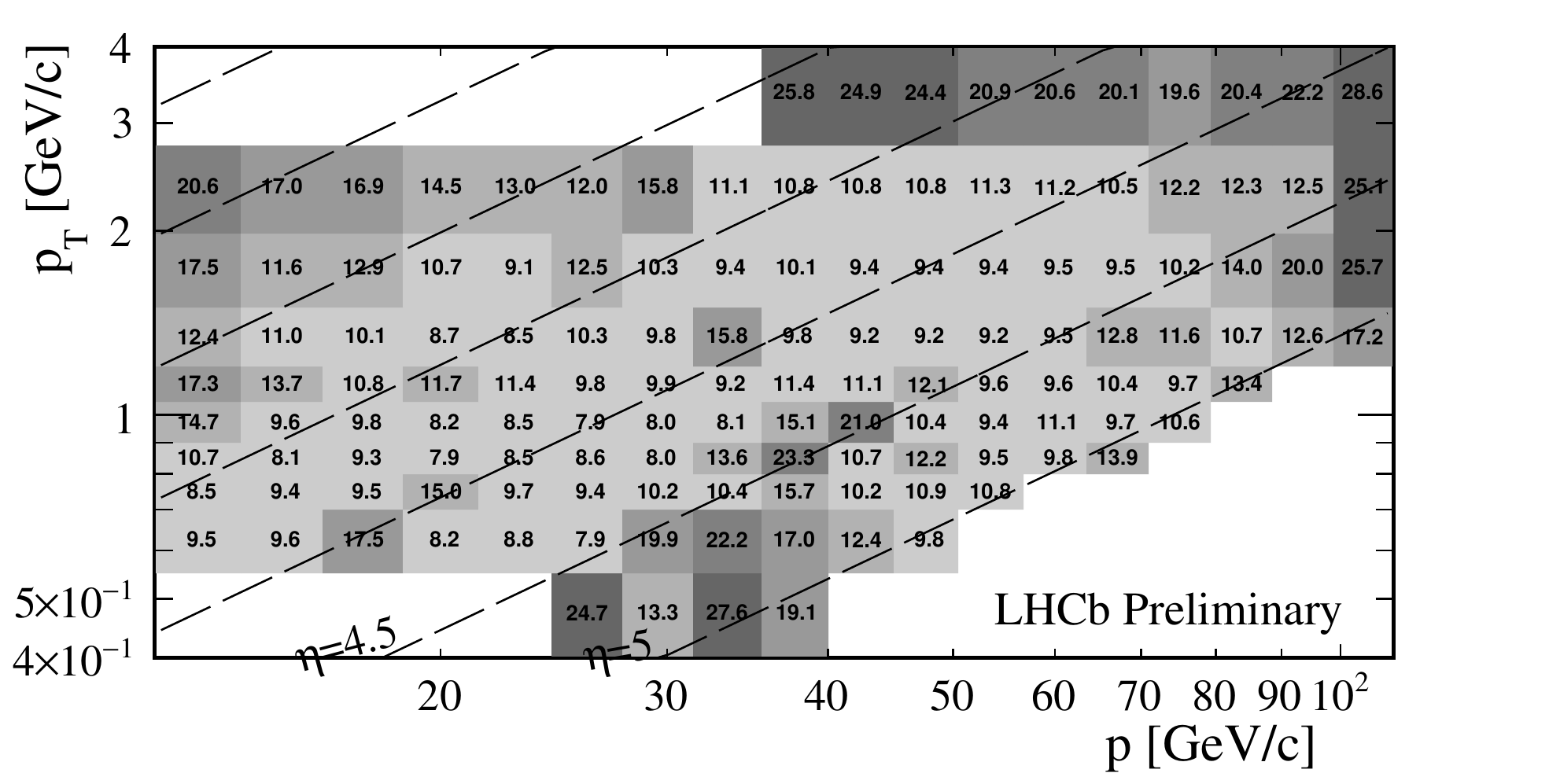}
\caption{Total relative uncertainty for the cross-section measurement
  in each kinematic bin, in per cent.}
\label{fig:errors}
\end{figure}

\subsection{Uncertainties}

Table~\ref{tab:syst} summarizes the uncertainties on the cross-section measurement.
The precision of the measurement is limited by the systematic uncertainty.
The largest uncertainty which is correlated among all kinematic bins is the aforementioned relative 6\%
on the normalization. 
The  uncorrelated uncertainty is dominated for most bins by the error 
on the \pbar fraction from the PID analysis.  Large relative uncertainties, up to 26\%, affect the
bins at the borders of the detector acceptance and, for the intermediate momentum
region, in the transition region between the two RICH detectors, at $\eta\sim 4.4$.
For the other regions, the accuracy is typically a few per cent.
The relative total uncertainty in each bin is illustrated in
Fig.~\ref{fig:errors}. It amounts to a relative 10\% or less for most of 
the accessible \pt regions.

A major difference between the fixed target configuration and the
standard \pp data taking in \lhcb is the extension of the luminous
region. Particular care is devoted to evaluate the dependence on the
\pvz of the different experimental effects. 
The \pbar yield normalized to the electron yield 
is found to be independent of \pvz within the statistical uncertainty.
The stability of the result is also checked as a function of the
absolute time and of the time within the  LHC orbit, excluding unexpected
biases related to the beam time structure.

\subsection{Results}

\begin{figure}[h]
    \centering
    \includegraphics[width=.97\textwidth]{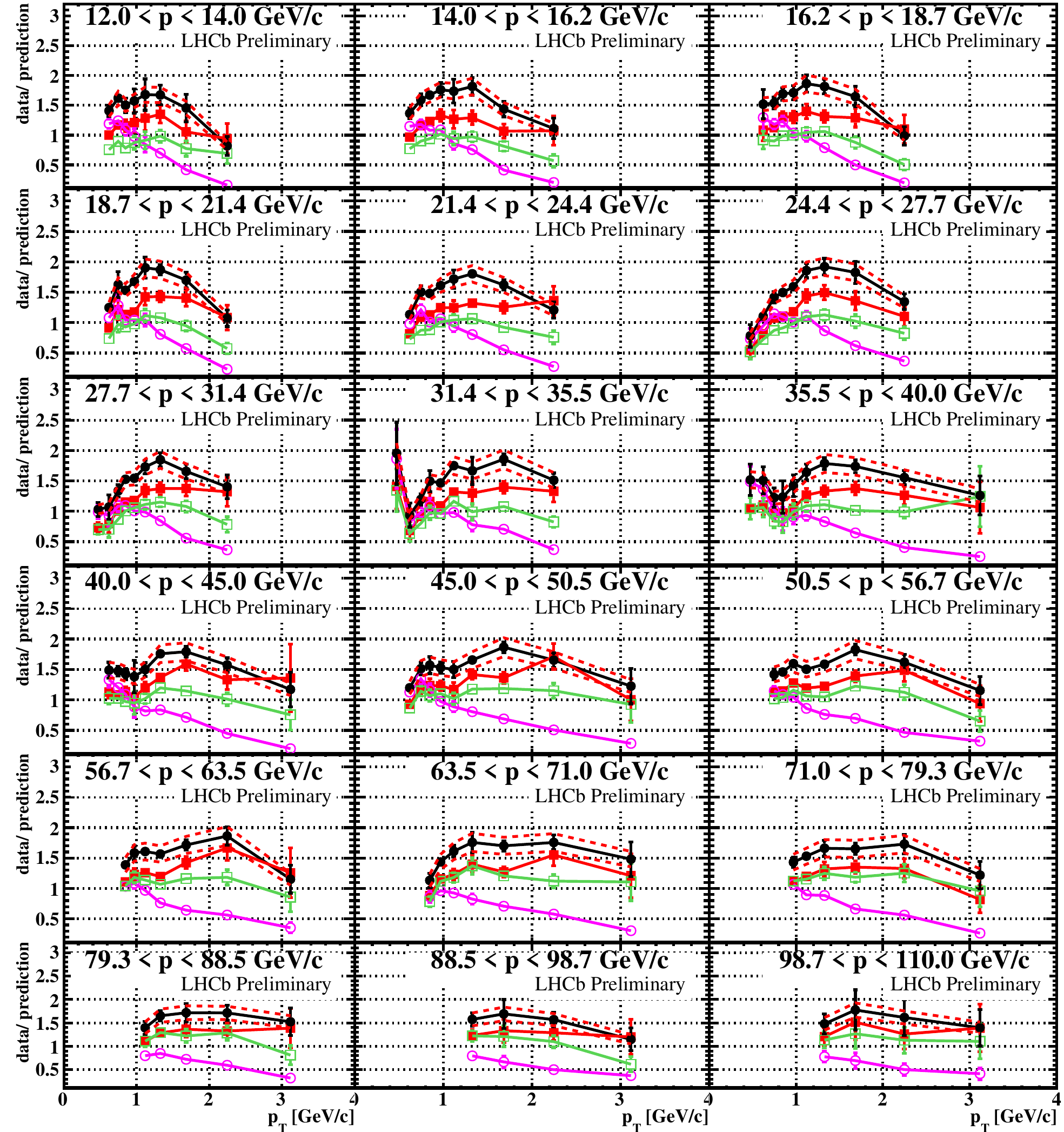}
    \caption{Result for the \pbar cross-section measurement, compared to absolute
predictions from different models. The plots show the ratio of data
over simulation as a function of \pt 
in the 18 momentum bins, for (black round closed symbols) EPOS LHC, (red squared closed symbols)
EPOS 1.99, (green squared open symbols) HIJING 1.38, and (violet round open symbols) QGSJETII-04.
The error bars represent the uncorrelated uncertainty 
for each  measurement. The additional
correlated uncertainty, shown only for EPOS LHC but also relevant 
to the other cases, is indicated by the red dashed lines.} 
    \label{fig:resultratio}
\end{figure}

The antiprotons candidates are counted from a sample of 33.7 million
selected \pHe collisions, from which  a sample of
1.4 million antiprotons is determined by the PID analysis. 
The double differential \pbar production cross-section
$d^2\sigma/dp\,dp_{\mathrm{ T}}$ is computed in
each kinematic bin after correcting for the reconstruction efficiency 
and the background.
The results are compared in Fig.~\ref{fig:resultratio} with the
predictions of the EPOS LHC~\cite{Pierog:2013ria} model, which is used
in the \lhcb simulation.  
The double
differential shape, notably the momentum spectrum, is found to be in good agreement with the
simulation, while the absolute production rate is larger on average by about a factor 1.5.
 The data are also compared with three
other models implemented in the CRMC~\cite{crmc} package v1.5.6: 
EPOS 1.99~\cite{Pierog:2009zt}, HIJING 1.38~\cite{Gyulassy:1994ew} and QGSJET II-04~\cite{Ostapchenko:2010vb}. 

The total inelastic cross-section is also determined  from the
measured total yield of recorded collisions within the
fiducial region. The PV reconstruction efficiency for inelastic
collisions is predicted, assuming the EPOS LHC model, to be $56.4 \pm
2.0\%$.
The result is
$ \sigma_{\rm inel}^{\rm LHCb} (\pHe, \sqsnn=110 \gev ) = (140 \pm 10) \text{~mb} $
which is larger than the EPOS LHC prediction
by a factor $1.19 \pm 0.08$,  implying that the measured \pbar multiplicity per inelastic collision 
is significantly larger in data. The multiplicity predicted by the pre-LHC version of EPOS 
is in better agreement with data. HIJING predicts a lower inelastic cross-section (100 mb),
while it reproduces well the measured absolute \pbar production cross-section values.
QGSJET matches the measured values  at very low \pt, while it exhibits a harder
\pt spectrum than data.
\section{Conclusions and Outlook}

The \lhcb experiment has recently opened the way to the use of the LHC
beams for fixed target physics. The first measurement of antimatter
production in \pHe collisions is one of the first results of this
novel program. Further details on this work can be found in the
related \lhcb conference contribution~\cite{LHCb-CONF-2017-002}.
The results are
expected to contribute to reduce the uncertainty on the prediction for
the secondary antiproton flux in cosmic rays. 
Further development of this study in the near future is foreseen, with
the inclusion of data collected at $\sqsnn=86.6$ GeV 
during November 2016, and the measurement of the contribution due to
hyperon decays. Several other measurements relevant to the modelling
of cosmic ray interactions will be possible in the near future: production of light
charged particles, including deuterons, photons and charmed particles
with different gas targets.

\section*{Acknowledgments}

We are indebted to our colleagues from the cosmic ray community,
O. Adriani, L. Bonechi, F. Donato and A. Tricomi, who suggested 
this measurement, and would like to thank F. Donato and T. Pierog for 
their advice.

\section*{References}
\newboolean{articletitles}
\setboolean{articletitles}{false}

\end{document}